\PassOptionsToPackage{table}{xcolor}
\documentclass[fleqn]{llncs}

\usepackage{cite}

\usepackage{amsmath} 
\usepackage{amssymb}
\usepackage{wasysym}
\usepackage{mathtools}
\usepackage{mathrsfs} 

\usepackage{verbatim}

\usepackage{ltltikz}

\usepackage{tikz}
\usetikzlibrary{arrows,automata,positioning,calc}
\usepackage{pgfplots}
\pgfplotsset{compat=1.14}

\usepackage{subfig}

\usepackage{arydshln}
\newcommand\donotshow[1]{}

\newcommand{\ldot}{\mathpunct{.}}

\newcommand{\set}[1]{{\{#1\}}}
\newcommand{\tuple}[1]{{\langle#1\rangle}}
\renewcommand{\models}{\vDash}

\newcommand{\bigo}{\mathcal{O}}
\newcommand{\pow}[1]{2^{#1}}
\renewcommand{\phi}{\varphi}

\newcommand{\card}[1]{{\left| {#1} \right|}} 

\newcommand{\graph}{\mathcal{G}}

\newcommand{\lang}{\mathcal{L}}

\newcommand{\ltl}{\text{LTL}}

\newcommand{\ltrue}{\mathit{true}}
\newcommand{\lfalse}{\mathit{false}}
\newcommand{\btrue}{\top}
\newcommand{\bfalse}{\bot}
\newcommand{\bimplies}{\rightarrow}

\newcommand{\nats}{\mathbb{N}}
\newcommand{\bool}{\mathbb{B}}

\newcommand{\fun}[2]{#1 \rightarrow #2}

\newcommand{\tsys}{\mathcal{T}}
\newcommand{\Paths}{\mathit{Paths}}

\newcommand{\ucw}{\mathcal{A}}
\newcommand{\sucw}{\mathscr{A}}

\newcommand{\rgeq}[1]{\vartriangleright_{#1}} 

\newcommand{\highlight}[1]{\colorbox{lightgray}{$\displaystyle #1$}}

\begin{document}

\title{Encodings of Bounded Synthesis\thanks{Supported by the European Research Council (ERC) Grant OSARES (No.\ 683300).}}
\author{Peter Faymonville\inst{1} \and Bernd Finkbeiner\inst{1} \and Markus~N.~Rabe\inst{2} \and Leander~Tentrup\inst{1}}
\institute{Saarland University \and University of California, Berkeley}

\maketitle

\begin{abstract}
The reactive synthesis problem is to compute a system satisfying a  given specification in temporal logic. 
Bounded synthesis is the approach to bound the maximum size of the system that we accept as a solution to the reactive synthesis problem. 
As a result, bounded synthesis is decidable whenever the corresponding verification problem is decidable, and can be applied in settings where classic synthesis fails, such as in the synthesis of distributed systems.
In this paper, we study the constraint solving problem behind bounded synthesis.
We consider different reductions of the bounded synthesis problem of linear-time temporal logic (LTL) to constraint systems given as boolean formulas (SAT), quantified boolean formulas (QBF), and dependency quantified boolean formulas (DQBF).
The reductions represent different trade-offs between conciseness and algorithmic efficiency.
In the SAT encoding, both inputs and states of the system are represented explicitly; in QBF, inputs are symbolic and states are explicit; in DQBF, both inputs and states are symbolic.
We evaluate the encodings systematically using benchmarks from the reactive synthesis competition (SYNTCOMP) and state-of-the-art solvers.
Our key, and perhaps surprising, empirical finding is that QBF clearly dominates both SAT and DQBF.
\end{abstract}

\section{Introduction}

There has been a recent surge of new algorithms and tools for the
synthesis of reactive systems from temporal specifications~\cite{conf/cav/JobstmannGWB07,conf/tacas/Ehlers11,conf/cav/BohyBFJR12,journals/sttt/FinkbeinerS13,conf/tacas/FinkbeinerT14}. Roughly,
these approaches can be classified into two categories:
\emph{game-based synthesis}~\cite{Buchi1969} translates the specification into an
deterministic automaton and subsequently determines the
winner in a game played on the state graph of this automaton; 
\emph{bounded synthesis}~\cite{conf/atva/ScheweF07a} constructs a constraint system that
characterizes all systems, up to a fixed bound on the size of the
system, that satisfy the specification.

The success of game-based synthesis is largely due to the fact that
it is often possible to represent and analyze the game arena symbolically,
in particular with BDDs (cf. \cite{conf/cav/JobstmannGWB07}). As a result, it has been possible to scale
synthesis to realistic benchmarks such as the AMBA bus protocol~\cite{conf/date/BloemGJPPW07}.
However, because the deterministic automaton often contains many more
states than are needed by the implementation, the synthesized
systems are often unnecessarily (and impractically) large (cf.~\cite{conf/vmcai/FinkbeinerJ12}).
This problem is addressed by bounded synthesis, where an
iteratively growing bound can ensure that the synthesized system is
actually the smallest possible realization of the
specification. However, bounded synthesis has not yet reached the same
scalability as game-based synthesis. A likely explanation for the
phenomenon is that the encoding of bounded synthesis into the
constraint system is ``less symbolic'' than the BDD-based
representation of the game arena.  Even though bounded synthesis tools typically
use powerful SMT solvers, a careful study of the standard encoding shows
that both the states of the synthesized system and its inputs are enumerated explicitly~\cite{journals/sttt/FinkbeinerS13}.

The question arises whether it is the encodings that need to be
improved, or whether the poor scalability points to a more fundamental
flaw in the underlying solver technology. To answer this question, we
reduce the bounded synthesis problem of linear-time temporal logic
(LTL) to constraint systems given as boolean formulas (SAT),
quantified boolean formulas (QBF), and dependency quantified boolean
formulas (DQBF).  The reductions are landmarks on the
spectrum of symbolic vs. explicit encodings.
All encodings represent the synthesized system in terms of its
transition function, which identifies the successor state in terms of
the current state and the input, and additionally in terms of an
output function, which identifies the output signals in terms of the
current state and the input, and annotation functions, which relate
the states of the system to the states of a universal automaton
representing the specification.

In the SAT encoding of the transition function, a separate boolean
variable is used for every combination of a source state, an input
signal, and a target state. The encoding is thus explicit in both the
state and the input.  In the QBF encoding, a universal quantification
over the inputs is added, so that the encoding becomes symbolic in the
inputs, while staying explicit in the states. Quantifying universally
over the states, just like over the input signals, is not possible in QBF
because the states occur twice in the transition function, as source
and as target. Separate quantifiers over sources and targets would
allow for models where, for example, the value of the output function
differs, even though both the source state and the input are the same.
In DQBF we can avoid such artifacts and obtain
a ``fully symbolic'' encoding in both the states and the input.

We evaluate the encodings systematically using benchmarks from the
reactive synthesis competition (SYNTCOMP) and state-of-the-art
solvers.  Our empirical finding is that QBF clearly dominates both SAT
and DQBF. While the dominance of QBF over SAT fits with our intuition
that a more symbolic encoding provides opportunities for optimization
in the solver, the dominance of QBF over DQBF is surprising. This
indicates that with the currently available solvers, the most symbolic
encoding (DQBF) is \emph{not} the best choice. Of course, with better
DQBF solvers, this may change: our benchmarks identify
opportunities for improvement for current DQBF solvers.

\subsubsection{Related Work.}

The game-based approach to the synthesis of reactive systems dates back to B\"uchi and Landweber's seminal 1969 paper~\cite{Buchi1969}.
Modern implementations of this approach exploit symbolic representations of the game arena, using BDDs (cf. \cite{conf/cav/JobstmannGWB07}) or decision procedures for the satisfiability of Boolean formulas (SAT-, QBF- and DQBF-solvers). We refer to \cite{conf/vmcai/BloemKS14} for a detailed comparison of the different methods.

Bounded synthesis belongs to the class of \emph{Safraless decision procedures}~\cite{conf/focs/KupfermanV05}. Safraless synthesis algorithms avoid the translation of the specification into an equivalent deterministic automaton via Safra's determinization procedure. Instead, the specification is first translated into an equivalent universal co-B\"uchi automaton, whose language is then approximated in a sequence of deterministic safety automata, obtained by bounding the number of visits to rejecting states~\cite{conf/atva/ScheweF07a}. Most synthesis tools for full LTL, including Unbeast~\cite{conf/tacas/Ehlers11}, and Acacia+~\cite{conf/cav/BohyBFJR12}, are based on this idea. 

Bounded synthesis~\cite{conf/atva/ScheweF07a} limits not only the number of visits to rejecting states, but also the number of states of the synthesized system itself.
As a result, the bounded synthesis problem can be represented as a decidable constraint system, even in settings where the classic synthesis problem is undecidable, such as
the synthesis of asynchronous and distributed systems (cf.~\cite{journals/sttt/FinkbeinerS13}). 
There have been several proposals for encodings of bounded synthesis.
The first encoding~\cite{conf/atva/ScheweF07a,Finkbeiner+Schewe/07/smt} was based on first-order logic modulo finite integer arithmetic. Improvements to the original encoding include the representation of transition systems that are not necessarily input-preserving, and, hence, often significantly smaller~\cite{journals/sttt/FinkbeinerS13}, the lazy generation of the constraints from model checking runs~\cite{conf/vmcai/FinkbeinerJ12}, and specification rewriting and modular solving~\cite{conf/vmcai/KhalimovJB13}. Recently, a SAT-based encoding was proposed~\cite{conf/rp/ShimakawaHY15}.ä
Another SAT-based encoding~\cite{conf/cav/FinkbeinerK16} bounds, in addition to the number of states, also the number of loops. A QBF-based encoding has been used in the related problem of solving Petri games~\cite{conf/birthday/Finkbeiner15}. Petri games can be used to solve certain distributed synthesis problems. They have, however, a significantly simpler winning condition than the games resulting from LTL specifications.

This paper presents the first encodings of bounded synthesis based on QBF and DQBF, and the first comprehensive evaluation of the spectrum of encodings from SAT to DQBF with state-of-the-art solvers. The encodings are significantly more concise than the previous SAT-based encodings and provide opportunities for solvers to exploit the symbolic representation of inputs and states. The empirical evidence shows that, with current solvers, the QBF encoding is superior to the SAT and DQBF encodings. A further contribution of the paper are the benchmarks themselves, which pinpoint opportunities for the improvement of the solvers, in particular for DQBF.

\section{Preliminaries}

Given a finite set of variables $V$, we identify boolean assignments $\alpha : V \to \bool$ as elements from the powerset of $V$, i.e., given $V$ and $\alpha$, then $\vec{v} = \set{ v \mid \alpha(v) = \btrue } \in 2^V$ is a representation of $\alpha$.
We use $\bool(V)$ to denote the set of propositional boolean formulas over the variables $V$. 

\paragraph{LTL.}

Linear-time temporal logic~(LTL) is the standard specification language for linear-time properties.
Let $\Sigma$ be a finite alphabet, i.e., a finite set of atomic propositions.
The grammar of $\ltl$ is given by
\begin{equation*}
  \varphi \Coloneqq
  p \mid
  \neg \varphi \mid
  \varphi \lor \psi \mid
  \varphi \land \psi \mid
  \Next \varphi \mid
  \varphi \Until \psi \mid
  \varphi \Release \psi \enspace,
\end{equation*}
where $p \in \Sigma$ is an atomic proposition.
The abbreviations $\ltrue \coloneqq p \lor \neg p$, $\lfalse \coloneqq \neg \ltrue$,  $\Eventually \varphi = \ltrue \Until \varphi$, and $\Globally \varphi = \lfalse \Release \varphi$ are defined as usual.
We assume standard semantics and write $\sigma \models \varphi$ if $\sigma \in (2^\Sigma)^\omega$ satisfies $\varphi$.
The language of $\varphi$, written $\lang(\varphi)$, is the set of $\omega$-words that satisfy $\varphi$.

\paragraph{Automata.}

A universal co-B\"uchi automaton $\ucw$ over finite alphabet $\Sigma$ is a tuple $\tuple{Q,q_0,\delta,F}$, where
$Q$ is a finite set of states,
$q_0 \in Q$ the designated initial state,
$\delta:Q \times 2^\Sigma \times Q$ is the transition relation, and
$F \subseteq Q$ is the set of rejecting states.
Given an infinite word $\sigma \in (2^\Sigma)^\omega$, a run of $\sigma$ on $\ucw$ is an infinite path $q_0 q_1 q_2 \dots \in Q^\omega$ where for all $i \geq 0$  it holds that $(q_i,\sigma_i,q_{i+1}) \in \delta$.
A run is accepting, if it contains only finitely many rejecting states.
$\ucw$ accepts a word $\sigma$, if \emph{all} runs of $\sigma$ on $\ucw$ are accepting.
The language of $\ucw$, written $\lang(\ucw)$, is the set $\set{\sigma \in (2^\Sigma)^\omega \mid \ucw \text{ accepts } \sigma}$.

We represent automata as directed graphs with vertex set $Q$ and a symbolic representation of the transition relation $\delta$ as propositional boolean formulas $\bool(\Sigma)$. The rejecting states in $F$ are marked by double lines.

\begin{lemma}
  Given an LTL formula $\varphi$, we can construct a universal co-B\"uchi automaton $\ucw_\varphi$ with $\bigo(2^\card{\varphi})$ states that accepts the language $\lang(\varphi)$.
\end{lemma}

\begin{figure}[t]
  \centering
  \subfloat[Universal co-B\"uchi automaton $\ucw_{\psi}$]{
    \begin{tikzpicture}[->,>=stealth',shorten >=1pt,auto,thick,scale=1,transform shape]
  \node[state] (init) {$q_0$};
  \node[state,accepting,below left=0 and 2 of init] (left) {$q_1$};
  \node[state,accepting,below right=0 and 2 of init] (right) {$q_2$};
  \node[state,accepting,below=0.5 of init] (error) {$q_e$};
  
  \draw (init) edge[<-] +(-0.75,0.75)
        (init) edge[loop above] node {$\btrue$} ()
        (init) edge node[swap] {$r_1$} (left)
        (init) edge node {$r_2$} (right)
        (init) edge node {$g_1 g_2$} (error)
        (left) edge[loop below] node {$\overline{g_1}$} ()
        (right) edge[loop below] node {$\overline{g_2}$} ()
        (error) edge[loop below] node {$\btrue$} ()
        ;
\end{tikzpicture}
    \label{fig:automaton-arbiter}
  }
  \hspace{1cm}
  \subfloat[Transition system $\tsys_{\mathit{arb}}$]{
    \begin{tikzpicture}[->,>=stealth',shorten >=1pt,auto,thick,scale=1,transform shape]
  \tikzstyle{state}=[draw,shape=rectangle,inner sep=7pt,align=center]

  \node[state] (t0) {$t_0$};
  \node[state,right=of init] (t1) {$t_1$};
  
  \draw (t0) edge[<-] +(-0.75,0.75)
        (t0) edge[bend left=20] node {$\btrue / g_1$} (t1)
        (t1) edge[bend left=20] node {$\btrue / g_2$} (t0)
        ;
  
  \node[below=1.8cm] {};  
  \node[right=0.9cm] {};  
  \node[left=0.9cm] {};
  
\end{tikzpicture}
    \label{fig:solution-arbiter}
  }
  \caption{A specification automaton over inputs $r_1, r_2$ and outputs $g_1, g_2$ and a realizing transition system.}
  \label{fig:arbiter-example}  
\end{figure}
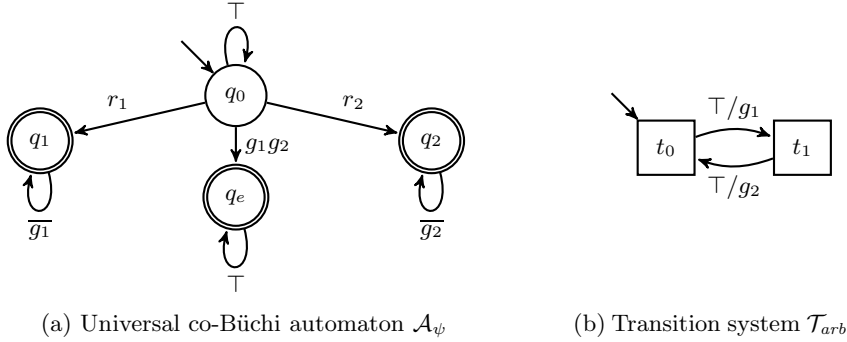

\begin{example}
  Consider the $\ltl$ formula $\psi = \Globally (r_1 \bimplies \Next\Eventually g_1) \land \Globally (r_2 \bimplies \Next\Eventually g_2) \land \Globally \neg(g_1 \land g_2)$.
  Whenever there is a request $r_i$, the corresponding grant $g_i$ must be set eventually.
  Further, it is disallowed to set both grants simultaneously.
  The universal co-B\"uchi automaton $\ucw_{\psi}$ that accepts the same language as $\psi$ is shown in Fig.~\ref{fig:arbiter-example}\subref{fig:automaton-arbiter}.
\end{example}

\paragraph{Transition Systems.}

In the following, we partition the set of atomic propositions into a set $I$ that contains propositions controllable by the environment and a set $O$ that contains propositions controllable by the system.
A transition system $\tsys$ is a tuple $\tuple{T,t_0,\tau}$ where
$T$ is a finite set of states,
$t_0 \in T$ is the designated initial state, and
$\tau : \fun{T \times \pow{I}}{\pow{O} \times T}$ is the transition function.
The transition function $\tau$ maps a state $t$ and a valuation of the inputs $\vec{i} \in 2^I$ to a valuation of the outputs, also called \emph{labeling}, and a next state $t'$.
If the labeling produced by $\tau(t,\vec{i})$ is independent of $\vec{i}$, we call $\tsys$ a state-labeled (or Moore) transition system and transition-labeled (or Mealy) otherwise.
Formally, $\tsys$ is a state-labeled transition system if, given a state $t \in T$ and any $\vec{i} \neq \vec{i'} \in \pow{I}$ with $\tau(t,\vec{i})=(\vec{o},\_)$ and $\tau(t,\vec{i'})=(\vec{o'},\_)$ it holds that $\vec{o} = \vec{o'}$.

Given an infinite word $\vec{i_0}\vec{i_1}\dots \in (2^I)^\omega$ over the inputs, $\tsys$ produces an infinite trace $(\set{t_0} \cup \vec{i}_0 \cup \vec{o_0}) (\set{t_1} \cup \vec{i}_1 \cup \vec{o_1}) \dots \in (\pow{T \cup I \cup O})^\omega$ where $\tau(t_j,\vec{i_j}) = (\vec{o_j}, t_{j+1})$ for every $j \geq 0$.
A path $w \in (\pow{I \cup O})^\omega$ is the projection of a trace to the atomic propositions.
We denote the set of all paths generated by a transition system $\tsys$ as $\Paths(\tsys)$.
A transition system realizes an $\ltl$ formula if $\Paths(\tsys) \subseteq \lang(\varphi)$.

\begin{example}
  Figure~\ref{fig:arbiter-example}\subref{fig:solution-arbiter} depicts the two-state (state-labeled) transition system $\tsys_{\mathit{arb}} = \tuple{\set{t_0,t_1},t_0,\tau}$ with $\tau(t_0,\vec{i}) = (\set{g_1},t_1)$ and $\tau(t_1,\vec{i}) = (\set{g_2},t_0)$ for every $\vec{i} \in 2^I$.
  The set of paths is $\Paths(\tsys) = (\set{g_1}\set{g_2})^\omega \cup (2^{\set{i_1,i_2}})^\omega$.
\end{example}

\section{Bounded Synthesis}

Bounded synthesis~\cite{journals/sttt/FinkbeinerS13} is a synthesis procedure for $\ltl$ specifications that produces size-optimal transition systems.
A given $\ltl$ formula $\varphi$ is translated into a universal co-B\"uchi automaton $\ucw$ that accepts the language $\lang(\varphi)$.
A transition system $\tsys$ realizes specification $\varphi$ if, and only if, every trace generated by $\tsys$ is in the language $\lang(\varphi)$.
$\tsys$ is accepted by $\ucw$ if every path of the unique run graph, that is the product of $\tsys$ and $\ucw$, has only finitely many visits to rejecting states.
This acceptance is witnessed by a bounded annotation on this product.

The bounded synthesis approach is to synthesize a transition system of bounded size $n$, by solving a constraint system that asserts the existence of a transition system and labeling function of $\tsys$ as well as a valid annotation. 
In this section we discuss how to construct a formula that represents that a \emph{given} annotation is correct. 
We will use this formula as a building block for different bounded synthesis constraint systems in Section~\ref{sec:encodings}.

The product of a transition system $\tsys = \tuple{T,t_0,\tau}$ and a universal co-B\"uchi automaton $\ucw = \tuple{Q,q_0,\delta,F}$ is a \emph{run graph} $\graph = \tuple{V,E}$, where
$V = T \times Q$ is the set of vertices and
$E \subseteq V \times V$ is the edge relation with
\begin{equation*}
  ((t,q),(t',q')) \in E \;\text{ iff }\; \exists \vec{i} \in 2^I \ldot \exists \vec{o} \in 2^O \ldot \tau(t,\vec{i}) = (\vec{o},t') \text{ and } (q,\vec{i} \cup \vec{o},q') \in \delta \enspace .
\end{equation*}
An annotation $\lambda : T \times Q \rightarrow \set{\bfalse} \cup \nats$ is a function that maps nodes from the run graph to either unreachable $\bfalse$ or a natural number $k$.
An annotation is valid if it satisfies the following conditions:
\begin{itemize}
  \item the pair of initial states $(t_0,q_0)$ is labeled by a natural number ($\lambda(t_0,q_0) \neq \bot$), and
  \item if a pair of states $(t,q)$ is annotated with a natural number ($\lambda(t,q) = k \neq \bot$) then for every $\vec{i} \in 2^I$ and $\vec{o} \in 2^O$ with $\tau(t,\vec{i}) = (\vec{o},t')$ and $(q,\vec{i} \cup \vec{o},q') \in \delta$, the successor pair $(t',q')$ is annotated with a greater number, which needs to be strictly greater if $q' \in F$ is rejecting. That is, $\lambda(t',q') \rgeq{q'} k$ where $\rgeq{q'} {}\coloneqq{} >$ if $q' \in F$ and $\geq$ otherwise.
\end{itemize}

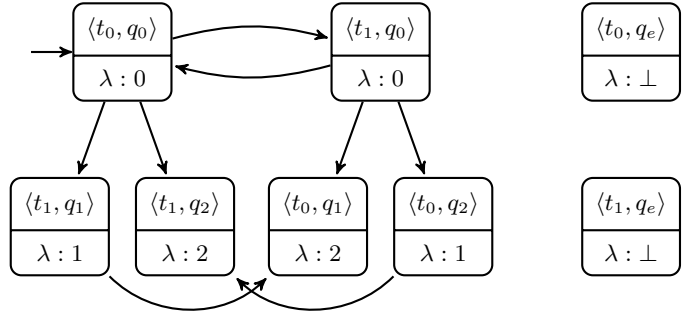
\begin{figure}[t]
  \centering
  \begin{tikzpicture}[->,>=stealth',shorten >=1pt,auto,thick,scale=1,transform shape]
  \tikzstyle{state}=[rounded corners,rectangle split, rectangle split parts=2, inner sep=5pt, draw]

  \node[state] (t0q0) {$\tuple{t_0,q_0}$ \nodepart{two} $\lambda: 0$};
  \node[state,right=2.1 of t0q0] (t1q0) {$\tuple{t_1, q_0}$ \nodepart{two} $\lambda: 0$};
  \node[state,below left=1 and -.5 of t0q0] (t1q1) {$\tuple{t_1, q_1}$ \nodepart{two} $\lambda: 1$};
  \node[state,below right=1 and -.5 of t0q0] (t1q2) {$\tuple{t_1, q_2}$ \nodepart{two} $\lambda: 2$};
  \node[state,below left=1 and -.5 of t1q0] (t0q1) {$\tuple{t_0, q_1}$ \nodepart{two} $\lambda: 2$};
  \node[state,below right=1 and -.5 of t1q0] (t0q2) {$\tuple{t_0, q_2}$ \nodepart{two} $\lambda: 1$};
  \node[state,right=2 of t1q0] (t0qe) {$\tuple{t_0,q_e}$ \nodepart{two} $\lambda: \bot$};
  \node[state,below=of t0qe] (t1qe) {$\tuple{t_1,q_e}$ \nodepart{two} $\lambda: \bot$};
  
  \draw (t0q0) edge[<-] +(-1.25,0)
        (t0q0) edge[bend left=15] (t1q0)
        (t1q0) edge[bend left=15] (t0q0)
        (t0q0) edge (t1q1)
        (t0q0) edge (t1q2)
        (t1q0) edge (t0q1)
        (t1q0) edge (t0q2)
        (t1q1) edge[bend right=45] (t0q1)
        (t0q2) edge[bend left=45] (t1q2)
        ;
\end{tikzpicture}
  \caption{Run graph  of the automaton $\ucw_{\psi}$ and the two-state transition system $\tsys_{\mathit{arb}}$ from the earlier example (Fig.~\ref{fig:arbiter-example}). The bottom node part displays a valid $\lambda$-annotation of the run graph.}
  \label{fig:run-graph-arbiter}
\end{figure}

\begin{example}
  Figure~\ref{fig:run-graph-arbiter} shows the run graph of $\tsys_{\mathit{arb}}$ and $\ucw_{\psi}$ from our earlier example (Fig.~\ref{fig:arbiter-example}).
  Additionally, a valid annotation $\lambda$ is provided at the second component of every node.
  One can verify that the annotation is correct by checking every edge individually.
  For example, the annotation has to increase from $\tuple{t_0,q_0} \rightarrow \tuple{t_1,q_2}$ and from $\tuple{t_0,q_2} \rightarrow \tuple{t_1,q_2}$ as $q_2$ is rejecting.
  As $\lambda(\tuple{t_0,q_0}) = 0$ and $\lambda(\tuple{t_0,q_2}) = 1$, it holds that $\lambda(\tuple{t_1,q_2})$ must be at least $2$.
\end{example}

Given $\tsys$, $\ucw$, and $\lambda$, we want to derive a propositional constraint that is satisfiable if, and only if, the annotation is valid.
First, by the characterization above, we know that we can verify the annotation by local checks, i.e., we have to consider only one step in the product graph.
To derive a propositional encoding, we encode $\tsys$, $\ucw$, and $\lambda$:
\begin{itemize}
  \item $\tsys = \tuple{T,t_0,\tau}$. We represent the transition function $\tau$ by one variable $o_{t,\vec{i}}$ for every output proposition $o \in O$ and one variable $\tau_{t,\vec{i},t'}$ representing a transition form $t$ to $t'$. Given $(t,t') \in T \times T$ and $\vec{i} \in 2^I$, it holds that (1) $\tau_{t,\vec{i},t'}$ is true if, and only if, $\tau(t,\vec{i}) = (\_,t')$, and (2) $o_{t,\vec{i}}$ is true if, and only if, $\tau(t,\vec{i}) = (\vec{o},\_)$ and $o \in \vec{o}$.
  \item $\ucw = \tuple{Q,q_0,\delta,F}$. We represent $\delta : (Q \times 2^{I \cup O} \times Q)$ as propositional formulas $\delta_{t,q,\vec{i},q'}$ over the output variables $o_{t,\vec{i}}$. 
      That is, an assignment $\vec{o}$ to the variables $o_{t,\vec{i}}$ satisfies $\delta_{t,q,\vec{i},q'}$ iff $(q,\vec{i} \cup \vec{o},q') \in \delta$.
  \item We first split the annotation $\lambda$ into two parts: The first part $\lambda^\bool : T \times Q \to \bool$ represents the reachability constraint and the second part $\lambda^\# : T \times Q \to \nats$ represents the bound. 
  For every $t\in T$ and $q\in Q$ we introduce variables $\lambda^\bool_{t,q}$ that we assign to be true iff the state pair is reachable from the initial state pair and a bit vector $\lambda^\#_{t,q}$ of length $\bigo(\log(\card{T} \cdot \card{Q}))$ that we assign the binary encoding of the value $\lambda(t,q)$. 
\end{itemize}
Using the variables $o_{t,\vec{i}}$, $\tau_{t,\vec{i},t'}$, $\lambda^\bool_{t,q}$, and $\lambda^\#_{t,q}$ (which have a unique assignment for a given $\mathcal T$, $\mathcal A$, and $\lambda$) as well as the propositional formulas $\delta_{t,q,\vec{i},q'}$, we construct a formula that represents that the annotation is valid:
\begin{align*} \label{eq:lambda-contraint-explicit}
  \hspace{-.3cm}
  \bigwedge_{q \in Q}
  \bigwedge_{t \in T}
    \left(\lambda^\bool_{t,q} \bimplies
    \bigwedge_{q' \in Q}
    \bigwedge_{\vec{i} \in 2^I}
    \left( \delta_{t,q,\vec{i},q'} \bimplies
      \bigwedge_{t' \in T} \left( \tau_{t,\vec{i},t'} \bimplies \lambda^\bool_{t',q'} \land \lambda^\#_{t',q'} \rgeq{q'} \lambda^\#_{t,q} \right)
    \right)
  \right)
\end{align*}

\begin{theorem}[\hspace{-.01pt}\cite{journals/sttt/FinkbeinerS13}]
  Given $\tsys$, $\ucw$, and an annotation $\lambda$.
  If the propositional encoding of $\tsys$, $\ucw$, and $\lambda$ satisfy the constraint system, then $\lambda$ is a valid annotation.
\end{theorem}

\section{Encodings}
\label{sec:encodings}

Using the constraints developed in the last section for checking the validity of a given annotation, we now consider the problem of finding a transition system with a valid annotation.

This section introduces four encodings, starting with the most explicit encoding and moving first to an input-symbolic variant, then to a input- and state-symbolic variant and then further to a ``fully symbolic'' variant which treats inputs, transition systems states and the specification automaton symbolically. The first encoding can be solved using a SAT solver, the second requires a QBF solver, and the remaining two encodings require a DQBF solver. 
We will indicate for each encoding the difficulty to switch from the decision variant of the problem (realizability) to the constructive variant of the problem (synthesis).

\subsection{SAT: The Basic Encoding} \label{sec:sat-encoding}

The \emph{basic encoding} of bounded synthesis follows almost immediately from the last section. 
Instead of checking that for given $\mathcal T$, $\mathcal A$, and $\lambda$, the unique assignment to the variables satisfies the formula, we existentially quantify over the variables to find an assignment. 
We only have to add constraints that assert that the reachability information, represented in the variables $\lambda^\bool_{t,q}$, is consistent, and that the transition relation, represented in the variables $\tau_{t,\vec{i},t'}$, provides at least one transition for every source state and every input. 
The consistency of the reachability annotation is given once we assert $\lambda^\bool_{t_0,q_0}$, as the formula itself asserts that the $\lambda^\bool_{t,q}$ annotations are consistent with the transition relation. 
\begin{flalign*}
 & \exists \set{\lambda^\bool_{t,q}, \lambda^\#_{t,q} \mid t \in T, q \in Q} &\\
 & \exists \set{\tau_{t,\vec{i},t'} \mid (t,t') \in T \times T, \vec{i} \in 2^I} &\\
 & \exists \set{o_{t,\vec{i}} \mid o \in O, t \in T, \vec{i} \in 2^I} &\\
 & \lambda^\bool_{t_0,q_0} \land \bigwedge_{t \in T} \bigwedge_{\vec{i} \in 2^I} \bigvee_{t' \in T} \tau_{t,\vec{i},t'} &\\
 & \bigwedge_{q \in Q}
  \bigwedge_{t \in T}
    \left(\lambda^\bool_{t,q} \bimplies
    \bigwedge_{q' \in Q}
    \bigwedge_{\vec{i} \in 2^I}
    \left( \delta_{t,q,\vec{i},q'} \bimplies
      \bigwedge_{t' \in T} \left( \tau_{t,\vec{i},t'} \bimplies \lambda^\bool_{t',q'} \land \lambda^\#_{t',q'} \rgeq{q'} \lambda^\#_{t,q} \right)
    \right)
  \right) &
\end{flalign*}

\begin{theorem}
  The size of the constraint system is in $\bigo(n m^2 \cdot 2^\card{I} \cdot (\card{\delta_{q,q'}} + n \log(nm)))$ and the number of variables is in $\bigo(n (m \log(nm) + 2^\card{I} \cdot ( \card{O} + n )))$, where $n = \card{T}$ and $m = \card{Q}$.
\end{theorem}

Since we only quantify existentially over propositional variables, the encoding can be solved by a SAT solver. The synthesized transition system can be directly extracted from the satisfying assignment of the solver. 
For each state and each input, there is at least one true variable, indicating a possible successor. The variables $o_{t,\vec{i}}$ indicate whether output $o$ is given at state $t$ for input $i$. 

\subsection{QBF: The Input-Symbolic Encoding} \label{sec:qbf-encoding}

One immediate drawback of the encoding above is the explicit handling of the inputs in the existential quantifiers representing the transition relation $\tau$ and the outputs $o$, which introduces several variables for each possible input $\vec{i} \in 2^I$.
This leads to a constraint system that is exponential in the number of inputs, both in the size of the constraints and in the number of variables. Also, since all variables are quantified on the same level, some of the inherent structure of the problem is lost and the solver will have to assign a value to each propositional variable, which may lead to non-minimal solutions of $\tau$ and $o$ due to unnecessary interdependencies.

By adding a universal quantification over the input variables, we obtain a quantified boolean formula (QBF) and avoid this exponential blow-up.
In this encoding, the variables representing the $\lambda$-annotation remain in the outer existential quantifier - they cannot depend on the input.
We then universally quantify over the valuations of the input propositions $I$ (interpreted as variables in this encoding) before we existentially quantify over the remaining variables. 

By the semantics of QBF, the innermost quantified variables, representing the transition function $\tau$ of $\tsys$, can be seen as boolean functions (Skolem functions) whose domain is the set of assignments to $I$.
Indicating the dependency on the inputs in the quantifier hierarchy, we can now drop the indices $\vec{i}$ from the variables $\tau_{t,\vec{i},t'}$ and $o_{t,\vec{i}}$. 
Further, we now represent $\delta : (Q \times 2^{I \cup O} \times Q)$ as propositional formulas $\delta_{t,q,q'}$ over the inputs $I$ and output variables $o_{\vec{t}}$ (which depend on $I$) with the following property: an assignment $\vec{i} \cup \vec{o}$ satisfies $\delta_{t,q,q'}$ iff $(q,\vec{i} \cup \vec{o},q') \in \delta$.
We obtain the following formula for the input-symbolic encoding.
(The gray box highlights the changes in the quantifier prefix compared to the previous encoding.)
\begin{flalign*}
&  \exists \set{\lambda^\bool_{t,q}, \lambda^\#_{t,q} \mid t \in T, q \in Q} &\\
&  \highlight{\forall I} &\\
&  \exists \set{\tau_{t,t'} \mid (t,t') \in T \times T} &\\
&  \exists \set{o_{t} \mid o \in O, t \in T} &\\
&  \lambda^\bool_{t_0,q_0} \land \bigwedge_{t \in T} \bigvee_{t' \in T} \tau_{t,t'} &\\
&  \bigwedge_{q \in Q}
  \bigwedge_{t \in T}
    \left(\lambda^\bool_{t,q} \bimplies
    \bigwedge_{q' \in Q}
    \left( \delta_{t,q,q'} \bimplies
      \bigwedge_{t' \in T} \left( \tau_{t,t'} \bimplies \lambda^\bool_{t',q'} \land \lambda^\#_{t',q'} \rgeq{q'} \lambda^\#_{t,q} \right)
    \right)
  \right) &
\end{flalign*}

\begin{theorem}
  Let $n = \card{T}$ and $m = \card{Q}$.
  The size of the input-symbolic constraint system is in $\bigo(nm^2 (\card{\delta_{q,q'}} + n \log(nm)))$.
  The number of existential and universal variables is in $\bigo(n (m \log(nm) + \card{O} + n))$ and $\bigo(\card{I})$, respectively.
\end{theorem}

The input-symbolic encoding is not only exponentially smaller (in $|I|$) than the basic encoding, but also enables the solver to exploit the dependency between $I$ and the transition function $\tau$. 
An additional property of this encoding that we use in the implementation is the following: If we fix the values of the $\lambda$-annotation, the resulting 2QBF query represents all transition systems that are possible with respect to the $\lambda$-annotation.
Since the outermost variables are existentially quantified, their assignments (in case the formula is satisfiable) can be extracted easily, even from non-certifying QBF solvers.
For synthesis, we thus employ a two-step approach.
We first solve the complete encoding and, if the formula was satisfiable, extract the assignment of the annotation variables $\lambda^\bool_{t,q}$, and $\lambda^\#_{t,q}$. 
In the second step we instantiate the formula by the satisfiable $\lambda$-annotation and solve the remaining formula with a certifying solver to generate boolean functions for the inner existential variables.
Those can be then be translated into a realizing transition system.

\subsection{DQBF/EPR: The State- and Input-Symbolic Encoding} \label{sec:symbolic-encoding}

The previous encoding shows how to describe the functional dependency between the inputs $I$ and the transition function $\tau$ and outputs $o$ as a quantifier alternation. 
The reactive synthesis problem, however, contains more functional dependencies that we can exploit.

In the following we describe an encoding that also treats the states of the system to generate symbolically. 
First, we change the definition of $T$ slightly.
Where before, $T$ was the set of states of the transition system, we now consider $T$ as the set of \emph{state bits} of the transition system.
Consequently, the state space of $\tsys$ is now $2^T$ and we consider the initial state to be the all-zero assignment to the variables $T$. 

Since all variables depend on the state, we no longer have propositional variables.
Instead, we quantify over the existence of boolean functions.
Candidate logics for solving this query are dependency-quantified boolean formulas~(DQBF) and the effective propositional fragment of first-order logic~(EPR).
While the existential quantification over functions is not immediately available in DQBF, we can encode them in a quadratic number of constraints, which is known as Ackermannization~\cite{conf/lpar/BruttomessoCFGSS06}. 
\begin{flalign*}
& \exists \set{\lambda^\bool_{q} \colon 2^T \to \bool, \lambda^\#_{q} \colon 2^T \to \bool^b \mid q \in Q} &\\
&  \exists \tau: 2^T \times 2^I \to 2^T &\\
&  \exists \set{ o \colon 2^T \times 2^I \to \bool \mid o \in O} &\\
&  \highlight{\forall I \ldot \forall T, T' \ldot } &\\
&  (T = 0 \bimplies \lambda^\bool_{q_0}(T)) &\\
&  \bigwedge_{q \in Q}
    \left(\lambda^\bool_{q}(T) \bimplies
    \bigwedge_{q' \in Q}
    \left( \delta_{q,q'} \land (\tau(T,I) \Rightarrow T') \bimplies \lambda^\bool_{q'}(T') \land \lambda^\#_{q'}(T') \rgeq{q'} \lambda^\#_{q}(T) \right)
  \right) &
\end{flalign*}

\begin{theorem}
  Let $n = \card{T}$ and $m = \card{Q}$.
  The size of the state-symbolic constraint system is in $\bigo(m^2 (\card{\delta_{q,q'}} + \log(nm)))$.
  The number of existential and universal variables is in $\bigo(n + m \log(nm) + \card{O})$ and $\bigo(n + \card{I})$, respectively.
\end{theorem}

\paragraph{Encoding the states of the specification automaton.}

The last dependency that we consider here is the dependency on the state space of the specification automaton.
As a precondition, we need a symbolic representation $\sucw = \tuple{Q,q_\text{init},\delta,q_\text{reject}}$ of a universal co-B\"uchi automaton over alphabet $I \cup O$, where
$Q$ is a set of variables whose valuations represent the state space,
$q_\text{init}\in \bool(Q)$ is a propositional formula representing the initial state,
$\delta\in\bool(Q,I \cup O,Q')$ is the transition relation ($\vec{q} \cup \vec{i} \cup \vec{o} \cup \vec{q'}$ satisfies $\delta$ iff $\vec{q} \xrightarrow{\vec{i} \cup \vec{o}} \vec{q'}$), and
$q_\text{reject} \in \bool(Q)$ is a formula representing the rejecting states.
\begin{flalign*}
& \exists \lambda^\bool \colon 2^T \times 2^Q \to \bool, \lambda^\# \colon 2^T \times 2^Q \to \bool^b &\\
&  \exists \tau: 2^T \times 2^I \to 2^T &\\
&  \exists \set{ o \colon 2^T \times 2^I \to \bool \mid o \in O} &\\
&  \highlight{\forall I \ldot \forall T, T' \ldot \forall Q, Q' \ldot } &\\
&  (t_\text{init} \land q_\text{init} \rightarrow \lambda^\bool(T, Q)) \land {} &\\
&   \left(\lambda^\bool(T, Q) \bimplies
    \left( \delta \land (\tau(T,I) \Rightarrow T') \bimplies \lambda^\bool(T',Q') \land \lambda^\#(T',Q') \rgeq{q'_\text{reject}} \lambda^\#(T, Q) \right)
  \right) &
\end{flalign*}

\begin{theorem}
  Let $n = \card{T}$ and $m = \card{Q}$.
  The size of the state-symbolic constraint system is in $\bigo(n + m + \card{\delta} + \log(nm))$.
  The number of existential and universal variables is in $\bigo(\log n + \card{O})$ and $\bigo(n + m + \card{I})$, respectively.
\end{theorem}

\subsection{Comparison}

Table~\ref{tbl:compare_encodings} compares the sizes of the encodings presented in this paper.
From the basic propositional encoding, we developed more symbolic encodings by making dependencies explicit and employing Boolean functions.
This conciseness, however, comes with the price of higher solving complexity.
In the following section we study this tradeoff empirical.

\begin{table}[t]
  \caption{The table compares the encodings with respect to the number of variables and the size of the constraint system. We indicate the number of states of the transition system and the automaton by $n$ and $m$, respectively.}
  \label{tbl:compare_encodings}
  \centering
  \begin{tabular}{l|lll}
    & \# existentials & \# universals \quad & constraint size \\ \hline 
    basic & $n (m \log(nm) + 2^\card{I} \cdot ( \card{O} + n ))$ & - & $n m^2 \cdot 2^\card{I}$ \\ &&& \quad$\cdot(\card{\delta_{q,q'}} + n \log(nm))$ \\
    input-symbolic & $n (m \log(nm) + \card{O} + n)$ & $\card{I}$ & $nm^2 (\card{\delta_{q,q'}} + n \log(nm))$ \\
    state-symbolic & $n + m \log(nm) + \card{O}$ & $n + \card{I}$ & $m^2 (\card{\delta_{q,q'}} + \log(nm))$ \\
    symbolic & $\log n + \card{O}$ & $n + m + \card{I}$ & $n + m + \card{\delta} + \log(nm)$ \\
  \end{tabular}
\end{table}

\section{Experimental Evaluation}

\begin{table}[t]
  \caption{Implementation matrix}
  \label{tbl:implementation-matrix}
  \centering
  \begin{tabular}{lllll}
    & basic \qquad\qquad & input-symbolic~ & state-symbolic~ & symbolic \\ \hline
    fragment & SAT & QBF & DQBF/EPR & DQBF/EPR \\
    Mealy/Moore & $\newmoon$/$\newmoon$ & $\newmoon$/$\newmoon$ & $\newmoon$/$\newmoon$ & $\newmoon$/$\newmoon$ \\
    solution extraction & $\newmoon$ & $\newmoon$ & $\fullmoon$ & $\fullmoon$ \\
  \end{tabular}
\end{table}

\subsection{Implementation}

We implemented the encodings described in this paper in a tool called \emph{BoSy}\footnote{The tool is available at~\url{https://react.uni-saarland.de/tools/bosy/}}.
The $\ltl$ to automaton conversion is provided by the tool ltl3ba~\cite{conf/tacas/BabiakKRS12}.
We reduce the number of counters and their size by only keeping them for automaton states within a rejecting strongly connected component, as proposed in~\cite{conf/vmcai/KhalimovJB13}.
The tool searches for a system implementation and a counter-strategy for the environment in parallel. An exponential search strategy is employed for the bound on the size of the transition system. 
In synthesis mode, we apply as a post-processing step circuit minimization provided by ABC~\cite{conf/cav/BraytonM10}.

For solving the non-symbolic encoding, we translate the propositional query to the DIMACS file format and solve it using the CryptoMiniSat SAT solver in version 5.
The satisfying assignment is used to construct the realizing transition system.

The input-symbolic encoding is translated to the QDIMACS file format and is solved by a combination of the QBF preprocessor Bloqqer~\cite{conf/cade/BiereLS11} and QBF solver RAReQS~\cite{journals/ai/JanotaKMC16}.
The solution extraction is implemented in two steps.
For satisfiable queries, we first derive a top level ($\lambda$) assignment~\cite{conf/date/SeidlK14} and instantiate the QBF query using this assignment which results in a 2QBF query that represents transition systems that satisfy the specification.
This is then solved using a certifying QBF solver, such as QuAbS~\cite{journals/corr/Tentrup16}, CADET~\cite{conf/sat/RabeS16}, or CAQE~\cite{conf/fmcad/RabeT15}.
Among those, QuAbS performed best and was used in the evaluation.
The resulting resulting Skolem functions, represented as AIGER circuit, are transformed into a representation of the transition system.

The symbolic encodings are translated to DQDIMACS file format and solved by the DQBF solver iDQ~\cite{conf/sat/FrohlichKBV14}.
Due to limited solver support, we have not implemented solution extraction.

For comparison, we also implemented an SMT version using the classical encoding~\cite{journals/sttt/FinkbeinerS13}.
We also tested the state-symbolic and symbolic encoding with state-of-the art EPR solvers, but the solving times were not competitive. 
Table~\ref{tbl:implementation-matrix} gives an overview over the capabilities of the implemented encodings.

\begin{table}[t]

	\caption{Experimental results on selected scalable instances. Reported is the maximal parameter value $k$ for which the instance could be solved and the cumulative solving time $t$ (in seconds) up to this point.}	
	\vspace{6pt}
\renewcommand{\arraystretch}{1.3}
\begin{tabular}{l|c;{1pt/1pt}c|c;{1pt/1pt}c|c;{1pt/1pt}c|c;{1pt/1pt}c|c;{1pt/1pt}c|}
	& \multicolumn{2}{c|}{basic}  & \multicolumn{2}{c|}{input-sym} & \multicolumn{2}{c|}{state-sym}   & \multicolumn{2}{c|}{Acacia}& \multicolumn{2}{c|}{Party} \\ 
	instance	& max $k$ & sum $t$ & max $k$ & sum $t$ & max $k$ & sum $t$ & max $k$ & sum $t$ & max $k$ & sum $t$ \\ \hline \hline
	simple-arbiter		& 7 & 1008.7 & \cellcolor{gray!50} 8 & \cellcolor{gray!50}2.7 & 3 & 100.5 & \cellcolor{gray!50}8 & 59.2 & 6 & 902.7 \\ \hline 
	full-arbiter		& 4 & 2994.5 & 3 & 0.6 & 2 & 13.3 &  \cellcolor{gray!50}5 & \cellcolor{gray!50}2683.4 & 3 & 111.7  \\ \hline 
	roundrob-arbiter	& \cellcolor{gray!50}4 & 143.1 & \cellcolor{gray!50}4 & 227.0 & 2 & 11.0 & \cellcolor{gray!50}4 & 345.6 & \cellcolor{gray!50}4 & \cellcolor{gray!50}19.2 \\ \hline 
	loadfull			& 5 & 268.7 & \cellcolor{gray!50}8 & \cellcolor{gray!50}44.2 & 2 & 25.1 & 4 & 83.7 & 4 & 213.5 \\ \hline 
	prio-abiter		& 4 & 176.5 & 4 & 1.6 & 2 & 0.4 &  \cellcolor{gray!50}6 & \cellcolor{gray!50}701.2 & 3 & 69.0  \\ \hline 
	loadcomp			& 5 & 36.9 & \cellcolor{gray!50}6 & \cellcolor{gray!50}639.4 & 3 & 432.1 &  5 & 387.8 & 5 & 212.7  \\ \hline 
	genbuf				& 2 & 1840.3 & 2 & 2711.8 & 0 & -- & \cellcolor{gray!50}5 & \cellcolor{gray!50}159.3 & 0 & --  \\ \hline 
	generalized-buffer	& 2 & 2093.8 & 2 & 3542.8 & 0 & -- & \cellcolor{gray!50}6 & \cellcolor{gray!50}3194.8 & 2 & 792.5  \\ \hline 
	load-balancer		& 5 & 1148.8 & \cellcolor{gray!50}8 & \cellcolor{gray!50}83.2 & 2 & 75.3 & 5 & 270.8 & 0 & --  \\ \hline 
	detector			& 6 & 1769.0 & \cellcolor{gray!50}8 & 1010.7 & 3 & 239.4 &  \cellcolor{gray!50}8 & \cellcolor{gray!50}261.6 & 5 & 370.3  \\ \hline 
	
\end{tabular}
	\label{tab:results}
\end{table}

\subsection{Setup \& Benchmarks}

For our experiments, we used a machine with a $3.6\,\text{GHz}$ quad-core Intel Xeon processor and $32\,\text{GB}$ of memory.
The timeout and memout were set to $1$ hour and $8\,\text{GB}$, respectively.
We use the LTL benchmark sets from the latest reactive synthesis competition (SYNTCOMP~2016)~\cite{journals/corr/JacobsBBK0KKLNP16}. The benchmarks include a variety of arbiter specifications of increasing complexity, load balancers, buffers, detectors as well as benchmark suites from previously existing tools. Some of the benchmark classes are parameterized in the number of clients or masters, which allows scaling them for experimental purposes. In total, the realizability benchmark suite of SYNTCOMP~2016 consists of 195 benchmarks. We have additionally added six instances from scalable benchmark classes of this set to cover larger parameter values, resulting in a total size of 201 benchmarks for our benchmark set.

For comparison, we run the other two solves that participated in the SYNTCOMP 2016, that is Acacia~\cite{conf/cav/BohyBFJR12}, a game-based solver, and Party~\cite{conf/cav/KhalimovJB13}, a variant of the SMT bounded synthesis.

\subsection{Realizability}

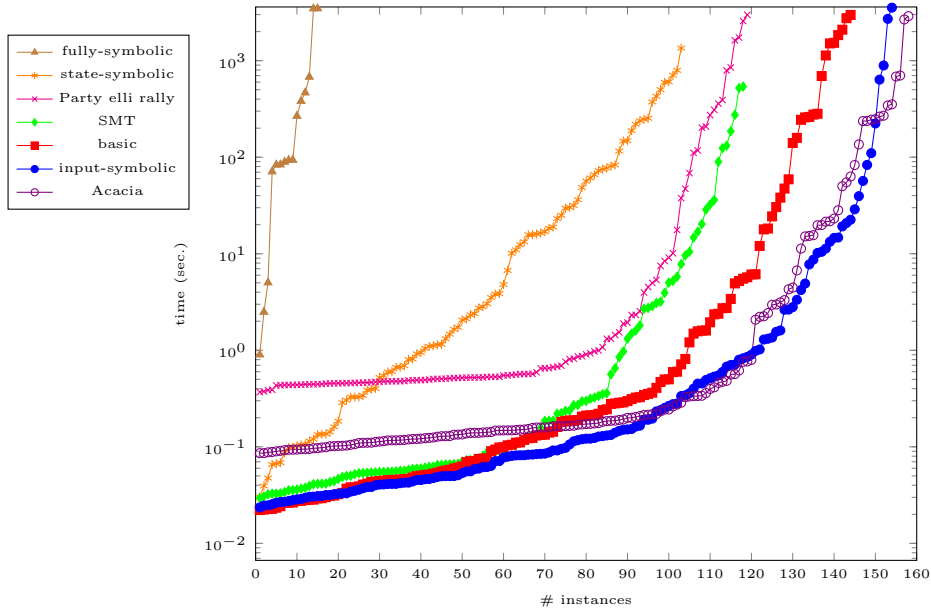
\begin{figure}[!tb]
  \centering
  \begin{tikzpicture}
    \begin{semilogyaxis}[xlabel=\# instances,ylabel=time (sec.),tiny,width=.85\columnwidth,mark size=1.7pt,ymin=0,ymax=3600,xmin=0,xmax=160,legend entries={fully-symbolic,state-symbolic,Party elli rally,SMT,basic,input-symbolic,Acacia},
        legend style={
          at={(-0.1,0.95)},
          anchor=north east}]
        ]]
      \addplot+[brown,solid,mark=triangle*] table {plots/bosy-paper_cactus_bosy-realizability-fully-symbolic-exponential-g0.dat};
      \addplot+[orange,solid,mark=asterisk] table {plots/bosy-paper_cactus_bosy-realizability-state-symbolic-exponential-g0.dat};
      \addplot+[magenta,mark=x] table {plots/bosy-paper_cactus_party_elli_rally-g0.dat};
      \addplot+[green,solid,mark=diamond*] table {plots/bosy-paper_cactus_bosy-realizability-smt-exponential-g0.dat};
      \addplot+[red,solid,mark=square*] table {plots/bosy-paper_cactus_bosy-realizability-non-symbolic-exponential-g0.dat};
      \addplot+[blue,solid,mark=*] table {plots/bosy-paper_cactus_bosy-realizability-input-symbolic-exponential-new-g0.dat};
      \addplot+[violet,solid,mark=o] table {plots/bosy-paper_cactus_acacia4aiger-g0.dat};

    \end{semilogyaxis}
  \end{tikzpicture}
  \caption{Number of solved instances within 1 hour among the 201 instances from SYNTCOMP~2016. The time axis has logarithmic scale.}
  \label{fig:cactus_solved_instance}
\end{figure}

In Table~\ref{tab:results}, we report results on realizability for all scalable instances from the aforementioned competition. We have omitted the results of our fully symbolic encoding from the table, since it could not solve a single instance of the selected benchmarks. The results from our own SMT encoding are also omitted, since they are very close to the results of the tool Party. Highlighted are those entries which reach the highest parameter value among the solvers and the best cumulative runtime within the class of instances.	
	
An overall comparison of all realizability solvers on the full benchmark set is provided in Figure~\ref{fig:cactus_solved_instance}. For the individual solvers, we track the number of instances solved by this solver within a a certain time bound.

\subsection{Synthesis}

To evaluate the different encodings in terms of their solutions to the synthesis problem and to compare with other competing tools, we measure the size of the provided solutions. In line with the rules of SYNTCOMP, the synthesized transition system is encoded as an AIGER circuit. The size of the result is measured in terms of the number of AND gates. In the comparisons, we only consider instances where both solvers in the comparison had a result. All resulting circuits have been minimized using ABC.

\begin{figure}[htb]
  \centering
  \begin{tikzpicture}
    \begin{loglogaxis}[
      only marks,
      axis equal,
      xmin=0,xmax=1409,
      ymin=0,ymax=1409,
      enlargelimits=false,
      width=.7\columnwidth,
      height=.7\columnwidth,
      xlabel=basic,
      xlabel style={
        at={(axis description cs:0.5,1)},
        anchor=south,
      },
      ylabel=input-symbolic,
      ylabel style={
        at={(axis description cs:1,0.5)},
        anchor=north,
      }]
      \addplot+[mark=+] table {./plots/bosy-paper_scatter_BoSy_synthesis_non-symbolic_exponential_vs_BoSy_synthesis_input-symbolic_exponential.dat};
      \draw[densely dashed,very thin] (0.1,0.1) -- (1409,1409);
      \draw[densely dashed,very thin] (0.1,0.01) -- (1409,140.9);
      \draw[densely dashed,very thin] (0.1,0.001) -- (1409,14.09);
    \end{loglogaxis}
  \end{tikzpicture}
  \caption{Scatter plot comparing the size of the synthesized strategies between the basic (Sec.~\ref{sec:sat-encoding}) and input-symbolic (Sec.~\ref{sec:qbf-encoding}) encoding. Both axes have logarithmic scale.}
  \label{fig:scatter_and_gates_nonsymb_encodings}
\end{figure}
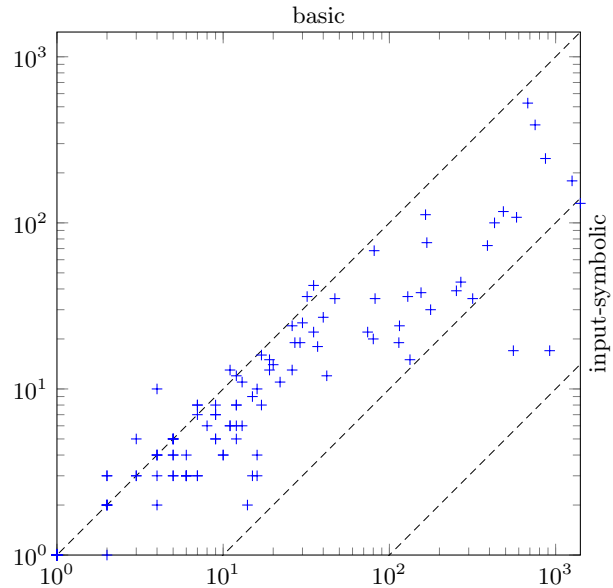

First, we compare in the scatter plot of Figure~\ref{fig:scatter_and_gates_nonsymb_encodings} the propositional, non-symbolic encoding to the input-symbolic encoding. Since most points are below the diagonal and are therefore smaller than their counterparts, the input-symbolic solutions are better in size compared to the non-symbolic encoding. 

\begin{figure}[!t]
  \centering
  \begin{tikzpicture}
    \begin{loglogaxis}[
      small,
      only marks,
      axis equal,
      xmin=0,xmax=103420,
      ymin=0,ymax=103420,
      enlargelimits=false,
      width=.5\columnwidth,
      height=.5\columnwidth,
      mark size=1.5pt,
      xlabel=Acacia,
      xlabel style={
        at={(axis description cs:0.5,1)},
        anchor=south,
      },
      ylabel=BoSy (input-symbolic),
      ylabel style={
        at={(axis description cs:1,0.5)},
        anchor=north,
      }]
      \addplot+[mark=+] table {./plots/bosy-paper_scatter_Acacia4Aiger-Synthesis_vs_BoSy_synthesis_input-symbolic_exponential.dat};
      \draw[densely dashed,very thin] (0.1,0.1) -- (103420,103420);
      \draw[densely dashed,very thin] (0.1,0.01) -- (103420,10342);
      \draw[densely dashed,very thin] (0.1,0.001) -- (103420,1034.2);
      \draw[densely dashed,very thin] (0.1,0.0001) -- (103420,103.42);
    \end{loglogaxis}
  \end{tikzpicture}
  \begin{tikzpicture}
    \begin{loglogaxis}[
      small,
      only marks,
      axis equal,
      xmin=0,xmax=1221,
      ymin=0,ymax=1221,
      enlargelimits=false,
      width=.5\columnwidth,
      height=.5\columnwidth,
      mark size=1.5pt,
      xlabel=Party elli rally,
      xlabel style={
        at={(axis description cs:0.5,1)},
        anchor=south,
      },
      ylabel=BoSy (input-symbolic),
      ylabel style={
        at={(axis description cs:1,0.5)},
        anchor=north,
      }]
      \addplot+[mark=+] table {./plots/bosy-paper_scatter_Party_elli_rally_vs_BoSy_synthesis_input-symbolic_exponential.dat};
      \draw[densely dashed,very thin] (0.1,0.1) -- (1221,1221);
      \draw[densely dashed,very thin] (0.1,0.01) -- (1221,122.1);
      \draw[densely dashed,very thin] (0.1,0.001) -- (1221,12.21);
    \end{loglogaxis}
  \end{tikzpicture}
  \caption{Scatter plot comparing the size of the synthesized strategies of BoSy, Acacia, and Party elli rally. Both axes have logarithmic scale.}
  \label{fig:scatter_and_gates}
\end{figure}
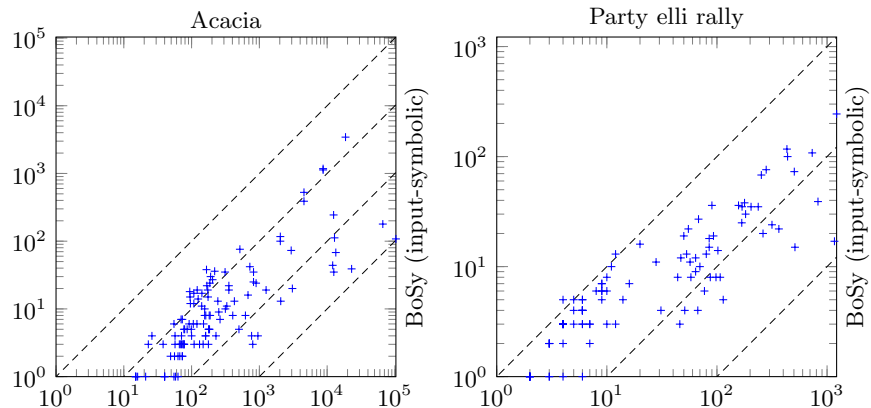

In Figure~\ref{fig:scatter_and_gates}, we compare our input-symbolic encoding against two competing tools. On the left, we observe that the solution sizes of our input-symbolic encoding are significantly better (observe the log-log scale) than the solutions provided by Acacia. The reason for the size difference is that the strategies of Acacia may depend on the current state of the specification automaton, as they are extracted from the resulting safety game. When comparing to the SMT-based Party tool, we again see a strict improvement in terms of strategy size, but not as significant as for Acacia.

We thus observe that the ability to universally quantify over the inputs and extract the transition system from the functional descriptions leads to advantages in terms of the size of the solution strategies.

\section{Conclusion}

We have revisited the bounded synthesis problem~\cite{journals/sttt/FinkbeinerS13} and presented alternative encodings into boolean formulas~(SAT), quantified boolean formulas~(QBF), and dependency-quantified boolean formulas~(DQBF).
Our evaluation shows that the QBF approach clearly dominates the SAT approach and the DQBF approach, and also previous approaches to bounded synthesis -- both in terms of the number of instances solved and in the size of the solutions. 
This demonstrates that, while modern QBF-solvers effectively exploit the input-symbolic representation, current DQBF solvers cannot yet take similar advantage of the state-symbolic representation. The benchmarks obtained from the encodings of bounded synthesis problems should therefore be useful in improving current solvers, in particular for DQBF.

\newpage
\bibliography{main}

\end{document}